# A Rule Based Expert System to Assess Coronary Artery Disease under Uncertainty

Sohrab Hossain[1*], Dhiman Sarma[2], Rana Joyti Chakma[2], Wahidul Alam[3], Mohammed Moshiul Hoque[4] and Iqbal H. Sarker[4]

[1] Department of Computer Science & Engineering,
East Delta University, Chittagong, Bangladesh.
[2] Department of Computer Science and Engineering,
Rangamati Science and Technology University, Rangamati, Bangladesh.
[3] Department of Computer Science and Engineering
University of Science and Technology Chittagong, Bangladesh.
[4] Department of Computer Science & Engineering,
Chittagong University of Engineering & Technology, Chittagong, Bangladesh.
{Correspondences: sohrab.h@eastdelta.edu.bd, iqbal@cuet.ac.bd}

**Abstract.** The coronary artery disease (CAD) occurs from narrowing and damaging of major blood vessels or arteries. It has become the most life threating disease in the world, especially in South Asian region. Its detection and treatment involve expensive medical facilities. The early detection of CAD, which is a major challenge, can minimize the patients' sufferings and expenses. The major challenge for CAD detection is incorporating numerous factors for detail analysis. The goal of this study is to propose a new Clinical Decision Support System (CDSS) which may assist doctors for analyzing numerous factors more accurately than the existing CDSSs. In this paper, a Rule Based Expert System (RBES) is proposed which involves five different Belief Rules, and can predict five different stages of CAD. The final output is produced by combining all BRBs and by using the Evidential Reasoning (ER). Performance evaluation is measured by calculating success rate, error rate, failure rate and false omission rate. The proposed RBES has higher success rate and false omission rate than other existing CDSSs.

## 1 Introduction

Coronary artery disease (CAD) is a condition when the coronary arteries become narrow or blocked. It is developed when bad cholesterols and plaques (fatty droplets) deposit inside the wall of arteries. The process is termed as atherosclerosis, means clogging of arteries, and reduces blood flow inside the heart muscle. Blood carries oxygen and essential nutrients to heart[1]. Lack of sufficient blood supply can cause angina (chest pain), and lead to heart attack by injuring heart muscle. The death toll due to heart disease is 16.3 million in America each year which has made it the leading cause of death in United States. According to American Heart Association (AHA), one person is suffered from heart attack in every 40 seconds. Having zero risk factor of heart disease, any male has 3.6% and any female has less than 1% chances of getting cardiovascular disease in his/her lifetime. Moreover, the chances are 37.5% and 18.3 respectively[2] for having 2 risk factors. In Bangladesh, CAD is responsible for 17% mortality rate[3]. Regular diagnostic approach of CAD relies on coronary angiogram test[4], echo-cardiogram ram (ECG)[5, 6], nuclear scan test and exercise stress test. ECG and exercise stress do not produce sustainable result for CAD prediction due to their non-invasiveness properties and numerous biases. Moreover, walking on a trade mill in stress test makes the patient discomfort heart function than normal condition. Now a days, Support Vector Machine (SVM) [7, 8] and Artificial Neural Network (ANN) [5, 8-17] based Clinical Decision Support Systems (CDSS) [18-20] are developed for CAD prediction. Unfortunately, SVM and ANN have no direct impact on the reasoning process due to their black-box- type modeling approaches. As a result, the degree of significance of individual factors cannot be resolved. So, human judgment and clinical data are both two essential factors for CAD diagnosis. For this purpose, CDSS combines both historical data and doctors' domain specific knowledge. But clinical data, like clinical domain knowledge, signs, and symptom, contain various uncertainties[21-23], and pose challenges for selecting domain knowledge to construct knowledge base. Moreover finding the reasoning under uncertainty requires excellent computational algorithm. To mitigate the challenges, researches introduced different CDSSs, based on fuzzy interface system and Bayesian interface system, which also have limitations [10, 24-26].In this paper, the proposed expert system can predict CAD by five classifications according to the severity. They are as follows:

Class A: (Normal or zero sign of heart disease).

Class B: (Unstable angina) - when new symptoms are introduced beside regular stable angina, and appears frequently (mostly when at rest), last long with more severity, and can lead to heart attack. It can be treated with oral medications (such as nitro-glycerine).

Class C: (Non-ST segment elevation myocardial infraction) - echocardiogram does not indicate the symptom of this type of myocardial infraction (MI) but chemical markers in the blood show the damage of heart muscles. The damage may not be significant and artery blockages are usually partial or temporary.

Class D: (ST segment elevation myocardial infraction) - this type of MI is occurred quickly due to sudden blockage by blood clogging. It can be detected by ECG and chemical markers in the blood, and causes damage of vast heart muscles.

Class E: (Silent ischemia)- Patient with heart disease can be suffered from sudden heart attack (called silent ischemia) without any prior or early warning, and the diabetic patients are common victims of this type[1].

## 2 Related Research

Researchers recently worked on machine learning and rule-based systems for different purposes [30, 31, 32, 33]. A number of researchersdeveloped belief-rule-based interference methodology by using evidential reasoning (RIMER) for CAD diagnosis [18, 27]. The RIMER process uses belief-rule-base for modeling clinical domain knowledge, and applies evidential reasoning approach for implementing reasoning. Studies show that RIMER based clinical decision support systems are highly efficient in supporting and interacting with clinical domain knowledge under uncertainty. In [28], Multi Criteria Decision Making Methods were presented for accessing CAD under uncertainty where presence and absence of CAD is predicted through using symptom and signs of CAD. But these approaches report neither the number of blocked arteries nor the significance of severity of the disease[8, 16, 26, 28, 29]. Weak parameters, like sign and symptoms, are used for predicting CAD as well as for predicting similar type of diseases like mitral regurgitation, dilated cardiomyopathy, congenital heart disease, hyper-tropic cardiomyopathy, myocardia infraction etc. Some researchers developed Medical Decision Support System (MDSS) to predict CAD. Other proposer polygenic risk scores (PRS), a nonlinear, for CAD prediction with accuracy of 0.92 under the receiver operating curve (AUC) [8].

Experimental analysis reveals that CAD diagnosis and its severity can be predicted significantly through clinical features along with pathological and demographic features[23, 25, 26, 28]. In this paper, we consider all these parameters, and proposed a cooperative-belief-rule based prototype (CDSS) to assist doctors for CAD analysis under uncertainty.

## 3 METHODOLOGY

### 3.1 Proposed Rule Based Expert System for CAD

In this paper, five separate BRBs are developed based on five distinct feature sets of patients such as: i) patients' pathological features, ii) patients' physiological features, iii) patients' demographic features, iv) patients' behavioral features, and iv) patients' non-modifiable risk factors. The BRBs are as follows:

$$D_A = f_A(S, P_A) \tag{1}$$

$$D_B = f_B(T, P_B) \tag{2}$$

$$D_C = f_C(X, P_C) \tag{3}$$

$$D_D = f_D(Y, P_D) \tag{4}$$

$$D_E = f_E(Z, P_E) \tag{5}$$

Here, S=$\{a_1, a_2, \ldots \ldots a_l\}$, T=$\{b_1, b_2, \ldots \ldots b_m\}$, X=$\{c_1, c_2, \ldots \ldots c_n\}$, Y=$\{d_1, d_2, \ldots \ldots d_o\}$, Z=$\{e_1, e_2, \ldots \ldots e_p\}$ represent the set of demographic, physiological, clinical, behavior-al, and Non-modifiable features respectively (where *l, m, n, o,* and *p* indicate the number of attributes for the five types of factors respectively).

Suppose that $P_A, P_B, P_C, P_D$ and $P_E$ are the corresponding vectors for the five BRBs, and $\omega=[\omega_1, \omega_2, \omega_3, \omega_4, \omega_5]$ represent the weight coefficients to the relative BRB where $f_A, f_B, f_C, f_D$ and $f_E$ are functions for demographic, physiological, clinical, behavioral, and non-modifiable factors respectively. To calculate individual matching degree for each rule, the following equation is used:

$$\alpha_{i,j} = \frac{u(A_{i,j+1}) - x_i}{u(A_{i,j+1}) - u(A_{i,j})} \tag{6}$$

Where $u$ is utility value, $a_{ij}$ is individual matching degree, $A_{ij}$ is $j^{th}$ referential value for $i^{th}$ attribute, and $x_i$ is the input for $i^{th}$ antecedent.

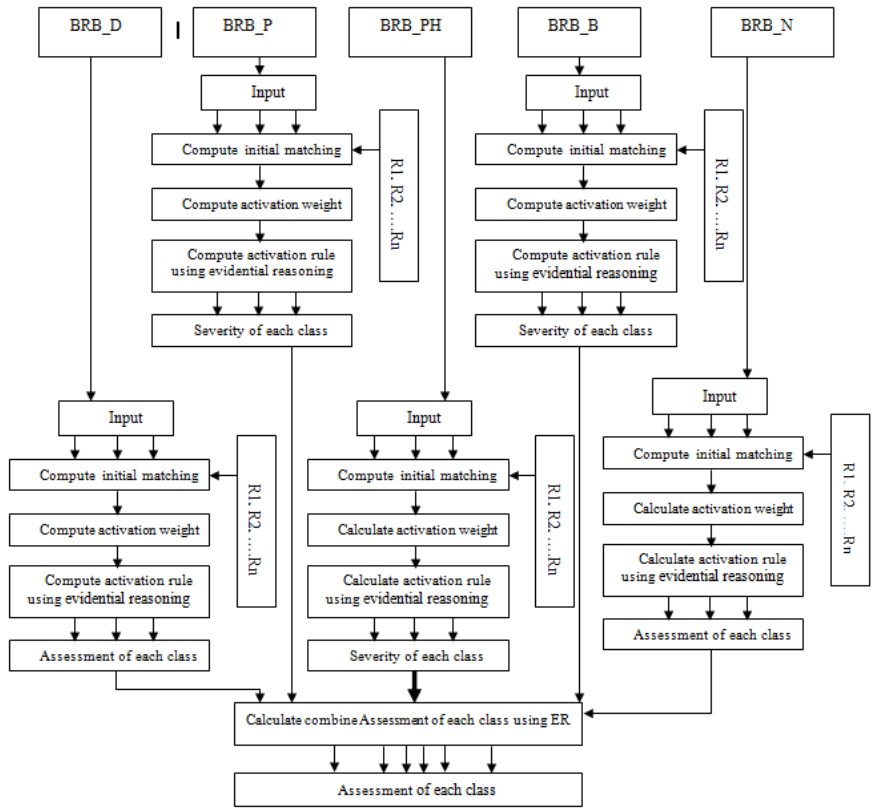

**Fig.1.** Rule Based Expert System to assess CAD

To calculate activated weight to each rule the following equation is used:

$$w_k = \frac{\theta_k \alpha_k}{\sum_{i=1}^{L} \theta_i \alpha_i} \tag{7}$$

Where $w_k$ is the $k^{th}$ rule's activation weight and $a_k$ is the interrelation between attributes. To calculate $a_k$, the following equations is used:

$$\alpha_k = \prod_{i=1}^{M} (\alpha_i^k)^{\bar{\delta}_i^k} \tag{8}$$

$$\bar{\delta}_i = \frac{\delta_i}{max_{i=1,\dots,M}(\delta_i)} \tag{9}$$

Where $\bar{\delta}_i$ is the normalized antecedent weight and $\alpha_i^k$ is the individual matching degrees for $i^{th}$ attribute. Five separate BRBs to predict CAD are : BRB_P, BRB_PH, BRB_D, BRB_B, and BRB_N. BRB_PH considers physiological factors like blood pressure and stress. BRB_P considers pathological factors like blood sugar level, low density lipoprotein, and triglyceride level. BRB_D considers factors like age and body mass index. BRB_B considers behavior factors like diet, smoking and physical activities. BRB_N considers non-modifiable risk factors like gender, family history, and residential Area.

### 3.2 Uncertainties in the Attribute

Attributes like blood pressure, stress, blood sugar level, triglyceride level, low density lipoprotein, age, body mass index, unhealthy diet, smoking, family history, and race are categorized into five classes, namely Physiological, Pathological, Demographical, Behavioral, and Non-modifiable risk factors. All the attributes have uncertainties at some level except gender attribute.

**Table 1.** Uncertainties in the attributes

| Attributes | Types of Uncertainty | Description |
|---|---|---|
| Blood Pressure, Stress, Blood sugar level, Triglyceride level, Low Density Lipoprotein | Impression | Information of the attributes are collected through medical instrumentations. The chances are high for storing wrong data in patient profile due to instrumental malfunctioning or operators' wrong procedural approaches. Moreover, the data for same patient may vary in different conditions. For example, blood pressure rises after some physical activities, blood sugar level falls after fasting and rises after meal. So, uncertainty in data may exist due to dataset ei different physical conditions or instrumental |
| | | malfunctions. |
| Age | Inconsistency | Old people may face hurdles of remembering their actual age. In under developed countries, young generation tend to hide actual age to apply for Government job. |
| Body mass index | Inconsistency | Body mass index is the ratio of weight and height. Patients' weight varies due to clothes, shoes, and after having meals. In this case, wrist measurement is more accurate which are maximum 35 inches for female and 40 inches for male. |
| Unhealthy diet | Vagueness, Incompleteness | Standard calorie demand varies from 1800 to 3000 calories and depends on height, weight, age, and physical activities. Exact amount of food consumption is very hard to measure, and accurate amount of calorie consumption may be unknown to patients. |
| Smoking | Incompleteness | Patients try to hide their smoking habit as they feel uncomfortable to disclose it in front of their family members. |
| Family history | Vagueness, Incompleteness | Most of the time, parents' heart disease information is not available. Even if some information available, the age, at which they got the disease, cannot be determined. |
| Race | Imprecision, Incompleteness | Usually patients use two or more addresses like present address, mailing address, and permanent address. |

## 3.3   Explanation of antecedent attributes

Five different types of attribute have been considered in this research. Explanations of numerical values of each attribute are as follows:

**Physiological factor.**

*Systolic Blood Pressure (SBP).*
The blood pressure which creates heart beats is known as systolic blood pressure. For SBP, five numerical points namely Usual, Elevated, Hypertension Stage 1, Hypertension Stage 2, Hypertension Stage 3 (hypertensive crisis) are considered and shown in table 2.

**Table 2.**Numerical value for blood pressure

| Terms | Numerical Values mm Hg (upper/lower) |
|---|---|
| Usual (U) | Less than 120/80 |
| Elevated (E) | 120/80-129/80 |
| Hypertension Stage 1(H1) | 130/89-139/89 |
| Hypertension Stage 2 (H2) | 140/90 or higher |
| Hypertensive Stage 3(H3) | Higher than 180/120 |
|  |  |

Here, the referential points can be presented as in equation (10).

$$PH1 \in \{U, E, H1, H2, H3\} \tag{10}$$

*Summed Stress Score (SSS):*
Intermediate risk of heart problem can be expressed in SSS score. SSS can be distributed into four referential points, namely regular, mildly irregular, moderately irregular and severely irregular are shown in table 3.

**Table 3.**Numericalvalue of Summed stress score

| Terms | Numerical Values |
|---|---|
| Regular (R) | <4 |
| Mildly irregular (M) | 4-8 |
| Moderately irregular (MI) | 9-13 |
| Severely irregular (S) | >13 |

The referential points can be presented as in equation (11).

$$PH2 \in \{R, M, MI, S\} \tag{11}$$

**Pathological factor.**

*Blood sugar level.* Blood sugar level is the level of sugar in blood. Five referential points are shown in table 4 and expressed by the equation (12).

**Table 4.** Numerical value of blood sugar level

| Terms | Amount shown  ( mg/dL) |
|---|---|
| Fasting (F) | Less than 100 |
| Before meal (B) | 70-130 |
| After meal (1-2 hours) (A) | Less than 180 |
| Before exercise (BE) | If taking insulin, at least 100 |
| Bed time (BT) | 100-140 |

$$P1 \in \{F, B, A, BE, BT\} \qquad (12)$$

ii)Triglyceride level

Triglyceride level measures triglycerides amount in blood. Four referential points are described in table 5 and equation (13).

**Table 5.** Numerical value of triglyceride level

| Terms | Triglyceride Level (mg/dL) |
|---|---|
| Healthy  (H) | Below 150 |
| Marginal high (BH) | 150-199 |
| High (H) | 200-499 |
| Extremely high (EH) | 500 and above |

$$P2 \in \{H,BH,H,EH\} \qquad (13)$$

*Low Density Lipoprotein(LDL).*

It contains both lipid and protein, and carries cholesterol to body tissues. Five referential values related to LDL are shown in table 6 and equation (14).

**Table 6.** Numerical value of low density lipoprotein

| Terms | Numerical Values ( mg/dL) |
|---|---|
| Good (D) | Less than 100; below 70 if CAD is present |
| Moderately elevated (NO) | 100-139 |
| Near to high (BH) | 140-159 |
| High (H) | 160-189 |
| Extremely high (VH) | 190 or Above |

$$P3 \in \{D,NO,BH,H,VH\} \qquad (14)$$

**Demographic factor.**

*Age.*

Older people are more likely to be victims of coronary artery disease, specially, after age of 65 years. Usually, older the age, higher the chance of getting CAD.

**Table 7.** Numerical values for age

| Terms | Numerical Values (age) |
|-------|------------------------|
| Young (Y) | <35 |
| Middle Age (M) | 35-49 |
| Old (O) | 50-65 |
| Extreme (E) | >65 |

Four referential values, namely, young (<35 years), mature (35-49 years), old (50-65), and extremely old (E), have been considered in the following equation from above table.

$$D1 \epsilon \{Y, M, O, E\} \qquad (15)$$

*Body mass index (BMI).*
Body mass index indicates the amount of body fat ratio and more effective for the age range from 18 to 65. It is the ratio of weight to height.

**Table 8.** Numerical value of BMI

| Terms | Numerical Points (BMI) |
|-------|------------------------|
| Healthy weight (H) | 18.5-24.9 |
| Overweight (O) | 25-29.9 |
| Obese (OB) | 30-39.9 |
| Morbid Obese (MO) | >=40 |

Four referential values, namely, healthy weight (18.5-24.9), overweight (25-29.9), obese (30-39.9), and morbid obese (>=40), have been considered in the following equation from the above table.

$$D2 \epsilon \{H, O, OB, MO\} \qquad (16)$$

**Behavior.**

*Unhealthy Diet.*

Mediterranean diet can reduce the risk of CAD by 30%. It is mainly plant based food and categorized into four sections shown in table 10 and expressed by equation (17).

**Table 9.** Numerical values of diet

| Terms | Numerical Points (Calories / day) |
|-------|-----------------------------------|
| Low (L) | <1800 |
| Healthy (H) | 1800–2200 |
| Moderate (M) | 2200 – 2800 |
| Eating Disorder (ED) | >2800 |
| | |

$$B1 \epsilon \{L, H, M, ED\} \qquad (17)$$

*Smoking.*

Smoking or exposers to smoke have high risk of CAD. Smoking is categorized into four sections, and shown in table 10 and expressed by equation (18).

**Table 10.** Numerical values for smoking

| Terms | Numerical Points (Cigarettes / day) |
|---|---|
| Non-Smoker (NS) | 0 |
| Smoker (S) | 1-5 |
| Moderate Smoker (MS) | 6-20 |
| Chain Smoker (CS) | >20 |

$$B2 \; \epsilon \; \{NS, S, MS, CS\} \qquad (18)$$

*Physical Activities.*

Inactive and less active people are in high risk to develop CAD. Physical activities are categorized into four sections which are shown in table 11 and expressed by equation (19).

**Table 11.** Numerical values for physical activities

| Terms | Numerical Points (Minutes / day) |
|---|---|
| Inactive (I) | 0-10 |
| Less Active (LA) | 11-20 |
| Active (A) | 21-30 |
| Very Active (VA) | >30 |

$$B3 \; \epsilon \; \{I, LA, A, VA\} \qquad (19)$$

**Non-modifiable risk factors.**

*Gender.*

Male has higher risk of CAD than female. In addition, male suffers from CAD in earlier age than female. But after age of 70 years, both male and female have similar chances of getting heart disease.

**Table 12.** Numerical values for gender

| Terms | Numerical Points |
|---|---|
| Male (M) | 0 |
| Female (F) | 1 |
| Other (O) | 2 |

$$N1 \; \epsilon \; \{M, F, O\} \qquad (20)$$

**Family history.**

If parents have histories of heart disease, children have high risk of developing CAD. The risk is even higher if parents have suffered before early 50 years of age.

The numerical points for the family history are represented by 0 (No history of parent's heart disease), 1 (History of parent's heart disease), and 2 (History of parent's heart disease before age of 50), and expressed in table 13 and by equation (21).

**Table 13.** Numerical values for family history

| Terms | Numerical Points |
|---|---|
| Low (L) | 0 |
| Medium (M) | 1 |
| High (H) | 2 |

$$N2 \ \epsilon \ \{L, M, H\} \tag{21}$$

**Residential Area.**

People from mega cities are more prone to CAD. This is because of higher rate of diabetes and obesity. On the other hand, people from hill track areas are less likely to develop heart disease. The numerical points for the residential areas are 0(Mega City), (Rural Area), and 2 (Hill track area), and expressed in table 14 and by equation (22).

**Table 14.** Numerical values for residential area

| Terms | Numerical Points |
|---|---|
| Low (L) | 0 |
| Medium (M) | 1 |
| High (H) | 2 |

$$N3 \ \epsilon \ \{L, M, H\} \tag{22}$$

### 3.4    Data set description

Dataset was collected from National Heart Foundation, Chittagong, Bangladesh with proper authorization. A summary of patients' data is given in Table.

**Table 15.** Summary of patients' data

| Patient Information | Number |
|---|---|
| Total patients | 1100 |
| Age intervalin year | 30-95 |
| Average age | 59 |
| Ratio (Male: Female) | 711:389 |
| Class A ( Normal) | 450 |
| Class B (Unstable angina) | 300 |
| Class C (Non-ST segment) | 287 |
| Class D (ST segment) | 43 |
| Class E (silent Ischemia) | 20 |

## 4 Result and Discussion

In binary diagnostic test, a positive or negative diagnosis is made for each patient. When the result of diagnosis is compared to the true condition, we find four possible outcomes: true positive, true negative, false positive, false negative.

### 4.1 Success Rate

Success rate is the ratio of correctly identified number of patients to total patients. Equation (23) is used to calculate the success rate and average success rate.

$$\text{Success Rate} = \frac{\text{Number of correctly identified patients}}{\text{Total patients}} \text{ X } 100\% \quad (23)$$

### 4.2 Error rate

Error rate is the ratio of incorrectly identified number of patients to total patients. Equation (24) is used to calculate the error rate and average error rate.

$$\text{Error Rate} = \frac{\text{Number of patients' incorrectly identified}}{\text{Total number of patients}} * 100\% \quad (24)$$

### 4.3 Failure rate

Failure rate is the ratio of number of non-recognized patients to total patients. Equation (25) is used to calculate the failure rate and average failure rate.

$$\text{Faliure Rate} = \frac{\text{Number of non recognised patients}}{\text{Total patients}} \text{ X } 100\% \quad (25)$$

### 4.4 False Omission Rate (FOR)

False omission rate is the ratio of number of patients identified to a class A to total number of patients belongs to a particular class except class A. Equation (26) is used to calculate the false omission rate and average false omission rate.

$$FOR = \frac{\text{Number of patients' identified to a class A}}{\text{Total number of patients belong to a particular class except class A}} \text{ X } 100\% \quad (26)$$

**Table 16.** Success rate, failure rate, error rate, false omission rate by expert system

| Class | Total No. of Patients (True Condition) | | | | | | Total Non-Recognized Patients | False Omission Rate (%) | Error rate (%) | Failure Rate (%) | Success Rate (%) |
|---|---|---|---|---|---|---|---|---|---|---|---|
| | | A | B | C | D | E | | | | | |
| | | | | | | | Total No. of Patients (Test Condition) | | | | |
| A | 450 | 405 | 7 | 3 | 0 | 25 | 10 | - | 07.78 | 10.00 | 90.00 |
| B | 300 | 15 | 268 | 10 | 0 | 07 | 10 | 05.00 | 08.33 | 10.67 | 89.33 |
| C | 287 | 1 | 07 | 270 | 10 | 0 | 6 | 00.35 | 03.83 | 05.92 | 94.08 |
| D | 43 | 2 | 03 | 02 | 36 | 0 | 0 | 04.65 | 16.27 | 16.27 | 83.72 |
| E | 20 | 05 | 0 | 0 | 0 | 10 | 05 | 25.00 | 25.00 | 50.00 | 50.00 |
| Average Rate (%) | | | | | | | | 03.54 | 08.81 | 11.10 | 89.90 |

Table 16 explains the results obtained by the equation number (23), (24), (25), and (26). Class A is considered as CAD negative patients and the remaining classes are CAD positive patients. It is observed that success rate of predicting class C type heart disease is the highest (94.08%) among five classes. On the other hand, class E prediction success rate is the lowest (50% only). Class E is very hard to predict as most of the time it does show any symptoms.

## 5      Conclusion

Heart disease is one of the major threats on public health and reason for the main cause of death worldwide. Although numerous researches are carried out in this area, still there are challenges to diagnose CAD for treatment. In this paper, the proposed expert system results an average accuracy rate of 89.90% which is the highest among other existing CDSS. The average false omission rate (3.54%) is also the lowest in this system than that of other CDSS. Our test results satisfy one of the main goals of this research. Average failure rate (11.10%) and average error rate (8.81%) are also remain as marginal. Class E (silent Ischemia) success rate is the lowest among all classes. The reason is that Class E occurs suddenly without showing any warning signs of heart problems. It was noted that Class E is common to the people with diabetes. It requires further research work to investigate whether or not diabetes influences the success rate in Class E type patients. Apart from this, our research concludes that RBES has higher success rate and false omission rate than other existing CDSS.